\documentclass{aa}
\usepackage{epsfig}
\usepackage{graphics}
\usepackage{subfigure}
\usepackage{natbib}
\usepackage{amsmath}
\usepackage{enumerate}

\begin{document}

\title{Neutrinos from collapsars}

\author{F. L. Vieyro\inst{1,2,3}, G. E. Romero\inst{1,2}, and O. L. G. Peres\inst{3,4}}
  
\institute{Instituto Argentino de Radioastronom\'{\i}a (IAR, CCT La Plata, CONICET), C.C.5, (1984) Villa Elisa, Buenos Aires, Argentina \and Facultad de Ciencias Astron\'omicas y Geof\'{\i}sicas, Universidad Nacional de La Plata, Paseo del Bosque s/n, 1900, La Plata, Argentina \and Instituto de F\'{\i}sica Gleb Wataghin, Universidade Estadual de Campinas, 13083-970 Campinas, SP, Brazil \and Abdus Salam International Centre for Theoretical Physics, ICTP, I-34010, Trieste, Italy}

\offprints{F. L. Vieyro \\ \email{fvieyro@iar-conicet.gov.ar}}

\titlerunning{Neutrinos from collapsars}

\authorrunning{Vieyro, et al.}

\abstract
{Long gamma-ray bursts (GRBs) are associated with the gravitational collapse of very massive stars. The central engine of a GRB can collimate relativistic jets that propagate inside the stellar envelope. The shock waves produced when the jet disrupts the stellar surface are capable of accelerating particles up to very high energies. }
{If the jet has hadronic content, neutrinos will be produced via charged pion decays. The main goal of this work is to estimate the neutrino emission produced in the region close to the surface of the star, taking pion and muon cooling into account, along with subtle effects arising from neutrino production in a highly magnetized medium. }
{We estimate the maximum energies of the different kinds of particles and solve the coupled transport equations for each species. Once the particle distributions are known, we calculate the intensity of neutrinos. We study the different effects on the neutrinos that can change the relative weight of different flavors. In particular, we consider the effects of neutrino oscillations, and of neutrino spin precession caused by strong magnetic fields. }
{The expected neutrino signals from the shocks in the uncorking regions of Population III events is very weak, but the neutrino signal produced by Wolf-Rayet GRBs with $z<0.5$ is not far from the level of the atmospheric background.}
{The IceCube experiment does not have the sensitivity to detect neutrinos from the implosion of the earliest stars, but a number of high-energy neutrinos may be detected from nearby long GRBs. The cumulative signal should be detectable over several years  ($\sim 10$ yr) of integration with the full 86-string configuration. }

 
\maketitle

\section{Introduction}

Gamma-ray bursts (GRBs) are the most violent and energetic events in the universe. Since GRBs are extragalactic sources, the equivalent isotropic energy can be as high as $10^{51} - 10^{54}$ erg \citep{bloom2001}. The generally accepted picture is that GRBs occur when the bulk kinetic energy of an ultra-relativistic flow is converted to internal energy through shocks and then is radiated away by non-thermal processes (e.g., \citealt{zhangMeszaros2004}).

Short GRBs seem to be the result of the final merger of two compact objects, whereas long GRBs are probably associated with the gravitational collapse of very massive stars. These imploding stars are called \textsl{collapsars}. The detection of supernova explosions weeks after several bursts strongly supports the association of long GRBs with the deaths of massive stars (e.g., \citealt{galama1998,bloom1999,reichart1999,lazzati2001}). The collapse of the stellar core produces a black hole, which accretes material from the inner layers of the star. An ultra-dense magnetized accretion disk is formed during the accretion process. Part of the plasma that surrounds the black hole is ejected, most likely by the magnetic pressure, producing two relativistic jets. Each jet then pushes the stellar material outwards. The location of the exact region where the gamma rays are created is still under debate. 

The most discussed model for explaining the origin of the prompt gamma-ray emission is the internal shock model \citep{reesMeszaros1994}. In this model, the central engine produces collimated shells that collide, creating internal shocks. Particles are accelerated up to relativistic energies in these shocks by a Fermi I-type mechanism. However, standard versions of the internal shock model do not explain the origin of the magnetic fields needed to produce the synchrotron radiation observed in the fireball. In addition, the validity of the model has been recently compromised, because it presents many problems for reproducing the variety of lightcurves observed with \textit{Swift} and \textit{Fermi} satellites in past years \citep{piran2007,ackermann2010}. Then, alternative models proposed to explain the gamma-ray emission are being currently explored.

Among these new models, we can mention models where the jet is magnetically dominated; in this case, the magnetic field is dragged from the highly magnetized central engine to the surface of the star. Internal shocks cannot be produced in magnetically dominated environments, so, in this context, the particle acceleration may be caused by dissipation of the strong magnetic fields and fast reconnection \citep{woosley1993,komissarov2009}.

Independently of the nature of the internal mechanism, it is widely accepted that the prompt emission has a different origin from the afterglow emission. The latter is emitted at a much greater distance from the central engine, when the fireball is decelerated by its interaction with the interstellar medium. (This occurs at $r_{\rm{es}} \sim 10^{17}$ cm, whereas the prompt gamma-ray emission is produced at $r_{\rm{es}} \sim 10^{13}$ cm.)

Besides producing electromagnetic emission (gamma rays from the prompt phase and radiation at lower energies from the afterglow), GRBs can also be sources of three important non-electromagnetic signals: cosmic rays, neutrinos, and gravitational waves. It seems reasonable to assume that if the prompt gamma-ray radiation and the afterglows are generated by relativistic electrons accelerated in shocks, then the same shocks should also accelerate baryons \citep{zhangMeszaros2004}. These high-energy protons can produce neutrinos through $pp$ inelastic collisions and $p \gamma$ interactions.

Several works have been devoted to studying the neutrino generation in different scenarios of GRBs. Neutrinos with energies in the range PeV to EeV ($10^{15-18}$ eV) can be produced by interactions of protons accelerated in the external forward and reverse shocks with the interstellar medium and surroundings (e.g., \citealt{waxmanBahcall2000,dai2001,razzaque2004}). Multi-TeV neutrinos may be created in the external reverse and forward shock produced by the interaction of the jet with the stellar envelope \citep{meszarosWaxman2001,horiuchi2008}; this signal is of special interest, because it can occur even if the jet fails to emerge from the star. These are the so-called \textsl{choked} GRBs. Hadrons can also be accelerated in the internal shocks, and their interaction with the prompt gamma-ray field may then lead to the production of PeV neutrinos \citep{waxman1997,guetta2001}. 

The production of neutrinos has been studied in different models for the prompt emission of GRBs; for example,  \citet{gao2012} estimates the neutrino emission in the GeV energy range for magnetized GRBs, and very recently, \citet{gaoAsano2012} and \citet{murase2013} have studied the production of GeV neutrinos in outflows loaded with neutrons, in which nuclear reactions result in subphotospheric gamma rays that can explain the prompt emission. The reader is referred to \citep{zhangMeszaros2004} for a thorough discussion of the different scenarios for neutrino emission.

Current upper limits set by IceCube have already ruled out the validity of some of these models and their predictions \citep{desiati2012}. The upper limit obtained with the data collected with the 59-string configuration of IceCube is 3.7 times below some theoretical predictions. This overestimation of the neutrino fluxes may be the result of several simplifications in the treatment of physical processes. The effects of the magnetic field in the cooling of transient charged particles may explain in part the deficit of neutrinos from collapsars. Magnetic fields within the jets of collapsars can take values as high as $10^{7-8}$ G close to the surface of the star, so synchrotron losses cannot be considered negligible for any charged particle. In addition, the intense radiation, matter, and magnetic fields in these sources modify the particle distributions that give rise to neutrinos. To compute the neutrino fluxes more accurately it is necessary, then, to treat the transport of photon and particles self-consistently.

In this work we propose a novel scenario for producing TeV neutrinos: the lateral shocks formed when the jet emerges from the stellar envelope. As the jet passes through the star, the material that is not being swept backward to the cocoon is pushed aside. At the moment the confinement produced by the stellar pressure ends, the lateral motion initiates a strong shock wave that moves around the star \citep{zhangWoosley2003}.  

We have calculated the neutrino generation in the lateral shocked regions and in the jet close to the stellar surface. First, in Sect. \ref{basicModel} we describe the geometry of the system, together with the basic assumptions and values of the relevant parameters. In Sect. \ref{particleAcc} we estimate the maximum energies of the different particle species and solve the set of coupled transport equations. Once the particle distributions are known, we proceed, in Sect. \ref{neutrinos}, to calculate the intensity of neutrinos of all flavors for each specific model under consideration. We also include the effects of neutrino oscillations and neutrino spin precession, which can change the relative weight of different flavors. In Sect. \ref{detectability}, we study the detectability of these sources with IceCube. Finally, in Sect. \ref{popIII}, we apply the model to GRBs with progenitors of Population III stars.

\section{Basic model}\label{basicModel}

The presence of the spectral features typical of Wolf Rayet (WR) stars in the afterglow emission favors these stars as the progenitor candidates for collapsars (e.g., \citealt{piro2000,mirabal2002}). We consider here a progenitor star with a radius of $R_{*} = 10^{12}$ cm. The exact value of the mass of the star has no influence on our calculation, because it will only change the duration of the GRB.

We assume that the core collapse leaves behind a black hole of initial mass $M_{\rm{BH}} = 10 M_{\odot}$ surrounded by an accretion disk. A fraction of the accreted matter is ejected into a coupled of relativistic jets, with a power red of

\begin{equation}
L_{\rm{j}} = 2 q_{\rm{j}} \dot{M} c^2,
\end{equation} 

\noindent where $\dot{M}$ is the accretion rate, and $q_{\rm{j}}$  an adimensional parameter that represents the efficiency of the mechanism at extracting rotational energy from the disk and converting it into the kinetic energy of the jet. A typical duration of a long GRB is a few tens of seconds, and in that time several solar masses are accreted by the black hole \citep{zhangWoosley2003}. Then, we adopt $\dot{M} = 0.1 M_{\odot}$ s$^{-1}$ and $q_{\rm{j}} = 0.02$. These values yield a luminosity of $L_{\rm{j}} \sim 1.8 \times 10^{51}$ erg s$^{-1}$, in agreement with observations \citep{zhangMeszaros2004}.

The jet is assumed to be formed at a distance $z_{0} = 50 r_{\rm{g}} \sim 10^8$ cm from the black hole, where $ r_{\rm{g}} = GM_{\rm{BH}}/c^2$ is the gravitational radius. This value also corresponds to the radius where the He core ends for a massive star \citep{horiuchi2008}. The jet becomes relativistic when it leaves the He core since the stellar envelope density drops considerably. 

After the jet crosses the He core, its opening angle remains approximately constant. We adopt $\theta = 10^{-1}$, i.e., approximately $6^{\circ}$. 

Assuming a conical geometry for the jet and an opening angle $ <<1 $, the radius of the jet is a function of the radius of the distance to the black hole $z$,

\begin{equation}\label{eq:rJet}
r_{\rm{j}}(z) = \theta z \sim r_0 \Big( \frac{z}{z_0} \Big),
\end{equation}

\noindent where $r_0 = r_{\rm{j}}(z_0)$ is the jet radius at the injection point, with a value of  $r_{\rm{j}}(z_0) = 10^7$ cm. This value is close to the size of the ergosphere of a high-spin Kerr black hole of $10 M_{\odot}$ \citep{meszarosRees2010}. We consider that the parameters only depend on the coordinate $z$ and do not depend on $r_{\rm{j}}$.

Since a magnetically driven mechanism is expected to be responsible for the jet launching, the magnetic energy density at the base of the jet should be related to the bulk kinetic energy density. In particular, if we assume equipartition between these energy densities, the jet magnetic luminosity and the comoving magnetic field at the base of the jet follow the relation:

\begin{equation}
\frac{L_{\rm{j}}}{4\pi r_{\rm{j}}(z_{0})^2 c \Gamma^2} = \frac{B_0^2}{8 \pi},
\end{equation}

\noindent where $B_0 = B(z_0)$, and $\Gamma$ is the jet bulk Lorentz factor. For a jet with a constant opening angle, the magnetic field strength decreases as a function of $z$, according to

\begin{equation}\label{eq:magneticField}
B(z) = B_{0} \Big( \frac{z}{z_0} \Big)^m ,
\end{equation}

\noindent with $1 \leq m \leq 2$, and $m=1$ corresponding to the lab-frame transverse component of the magnetic field (e.g., \citealt{krolik1999}).

By the time the jet emerges from the star, the magnetic energy density $w_{\rm{mag}}$ has decreased considerably, and it is only a fraction of the kinetic energy density $u_{\rm{kin}}$.  Since $w_{\rm{mag}} / u_{\rm{kin}} << 1$ at $z=R_{*}$, shocks can develop in the jet at the surface of the star \citep{komissarov2007}. These shocks can accelerate particles up to relativistic energies through diffusive shock acceleration.

\begin{figure}
\begin{center}
  \includegraphics[width=.4\textwidth,keepaspectratio=true]{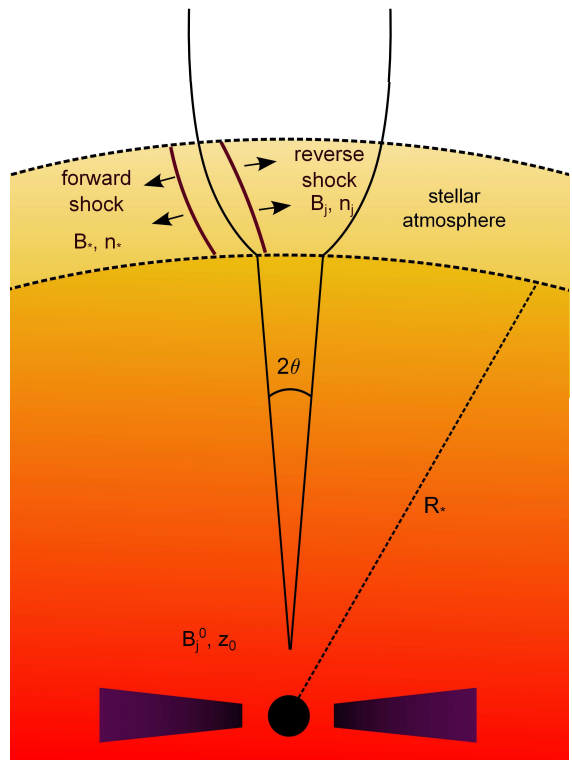}
  \caption{Schematic diagram of the jet model and the structure of double shock. }
  \label{fig:jet}
  \end{center}
\end{figure}

Numerical simulations of the collapsar model show that, at the moment the jet breaks the surface, it still has high internal energy. This causes the acceleration of the jet, which also expands abruptly near the stellar surface. Additionally, as the jet propagates inside the the star, it is collimated by the external pressure. Part of the material that is being pushed by the jet enters the working surface and moves backflows forming the cocoon. The remaining material spreads laterally when the jet emerges from the stellar surface, producing lateral forward shock waves that move around the star \citep{zhangWoosley2003}, and lateral reverse shocks that propagate inside the jet. To distinguish these shocks, we call them \textsl{forward shock} (FS) and \textsl{reverse shock} (RS). A schematic representation of the double shock structure, together with the geometry adopted for the jet, is shown in Fig. \ref{fig:jet}.

\subsection{Lateral reverse shock}

We represent the RS region as a cylinder with radius equal to the jet radius at $R_{*}$; this can be obtained from Eq. \ref{eq:rJet}, and results in $r_{\rm{j}} (R_{*}) = 10^{11}$ cm. The height of the cylinder is taken to be twice the radius. We consider a region small enough that the values of the parameters do not change considerably, so the one-zone approximation is valid.  In the comoving jet frame the particle density of the jet is given by

\begin{equation}\label{eq:njet}
n_{\rm{j}}(z)= \frac{L_{\rm{iso}}}{4\pi z^2 \Gamma_{\rm{j}}^2 m_{p} c^3}.
\end{equation}

\noindent This results in $n_{\rm{j}}(R_{*}) \sim 10^{16}$ cm$^{-3}$. 

For simplicity we consider a jet with a constant Lorentz factor. The inferred values for the Lorentz factor when the prompt gamma-ray emission is produced are in the range $100 < \Gamma < 10^3$ \citep{lithwick2001}. Here we adopt $\Gamma = 200$. 

The value of the magnetic field inside the jet at $z=R_{*}$ can be obtained from Eq. \ref{eq:magneticField}: $B_{\rm{RS}} \sim 10^7$ G. Because of the high value of the magnetic field, the synchrotron radiation by electrons would result in a very dense photon field in this region.

\subsection{Lateral forward shock}

The size of this region is taken equal to the rs region. As was pointed out previously, the progenitor of long GRBs are WR stars. These stars have powerful winds with a typical mass loss rate of $10^{-5}$ solar masses per year. Then, interaction with the matter field plays an important role in this region. We consider a particle density on the stellar surface of $n_{\rm{s}} = 2 \times 10^{13}$ cm$^{-3}$, which corresponds to a mass loss of $\sim 10^{-5} M_{\odot}$ yr$^{-1}$ with a terminal wind velocity of $\sim 1000$ km s$^{-1}$ \citep{zhangWoosley2003}.

Since this shock propagates on the stellar surface, the magnetic field is the one at the surface of the star. Here we adopt $B_{\rm{s}} = 100$ G, which is a value within the range of estimates for massive stars \citep{kholtyginline2011}. Given this relatively low value, synchrotron radiation would not provide a significant photon target field. Instead, the stellar photon field would be the most relevant target for IC scattering and photomeson production. We consider a WR star with an effective temperature of $45000$ K, which corresponds to a luminosity of $\sim 10^{39}$ erg s$^{-1}$ and a peak energy of $E_{\rm{p}} \sim 4$ eV \citep{sander2012}.  We use a Planck function to describe the nearly black-body emission of the star $N_{\rm{ph}}(E_{\rm{ph}})$ (in erg$^{-1}$ cm$^{-3}$), given by

\begin{equation}
N_{\rm{ph}}(E_{\rm{ph}}) = A_{\rm{ph}} \frac{ E_{\rm{ph}}^2} {\exp(E_{\rm{ph}}^2/ kT_{\rm{BB}}) - 1}.
\end{equation}

\noindent This is the photon density in the laboratory frame; to obtain the photon density in the shock co-moving frame, we use the standard Lorentz transformations.

\section{Particle distributions}\label{particleAcc}

\subsection{Particle acceleration and radiative losses}

Approximately 10\% of the energy goes to accelerate particles in the shock region, whereas the magnetic energy is $\sim 0.01L_{\rm{j}}$ or lower \citep{zhangMeszaros2004,panaitescu2001,panaitescu2002}. We consider then that 10\% of the energy of the jet is injected into relativistic particles, $L_{\rm{rel}}=q_{\rm{rel}}L_{\rm{j}}$, with $q_{\rm{rel}}=0.1$.

We adopt the power injected into leptons, $L_{e}$, to be a fraction of the power in protons, $L_{p}$, that is, $L_{e}=aL_{p}$. Recently, \citet{gaoMeszaros2013} has studied the consequences of the non-detection of neutrinos from the burst GRB 130427A. They obtained values for $a$ in the range $0.1-1$, so we adopt $a=0.1$.

For simplicity, we consider all the relevant parameters constant during most of the event. The values of these parameters are listed in Table \ref{table}. We study the radiative losses in both the FS and RS, and analyze the maximum energies that particles can achieve in each of these regions.

\begin{table}[h]
    \caption[]{Main parameters of the model.}
        \label{table}
        \centering
\begin{tabular}{ll}
\hline\hline %
Assumed parameters & Value\\ [0.01cm]
\hline
$M_{\rm{BH}}$:  black hole mass [$M_{\odot}$]                                                   & $10$\\
$\dot{M}$:      accretion rate [$M_{\odot}$ s$^{-1}$]             & $0.1$\\
$q_{\rm{j}}$:   launching efficiency                                                                     & $0.01$\\
$L_{\rm{j}}$:   jet luminosity [erg s$^{-1}$]                         & $10^{51}$\\
$\theta$:       jet opening angle                         & $0.1$\\
$\Gamma$:       jet Lorentz factor                        & $200$\\
$R_{*}$:        stellar radius [cm]                                                                                      & $10^{12}$\\
$z_{0}$:        jet injection radius [cm]                                                                & $10^{8}$\\
$q_{\rm{rel}}$:                                                 fraction of power injected in relativistic particles     & $0.1$\\
$a$:                                            lepton-to-hadron energy ratio                                                 & $0.1$\\
$\eta$:                                 acceleration efficiency                                                                         & $0.5$\\
$\alpha$:                       injection power-law index                                     & $2$      \\

\\
\hline\hline
Forward shock region parameters & Value \\ 
\hline
$n_{i}$:        plasma density [cm$^{-3}$]        & $\sim 100$\\
$B_{\rm{FS}}$:  magnetic field [G]                & $\sim 100$\\
\\
\hline\hline
Reverse shock region parameters & Value \\ 
\hline
$n_{i}$:   plasma density [cm$^{-3}$]        & $\sim 10^{16}$\\
$B_{\rm{RS}}$:  magnetic field [G]       & $\sim 10^7$\\

\hline  \\[0.005cm]
\end{tabular}   
                                                                                                                                                                        
\end{table} 

\begin{figure*}[t]%
\begin{center}
 \parbox{3in}{\epsfig{figure=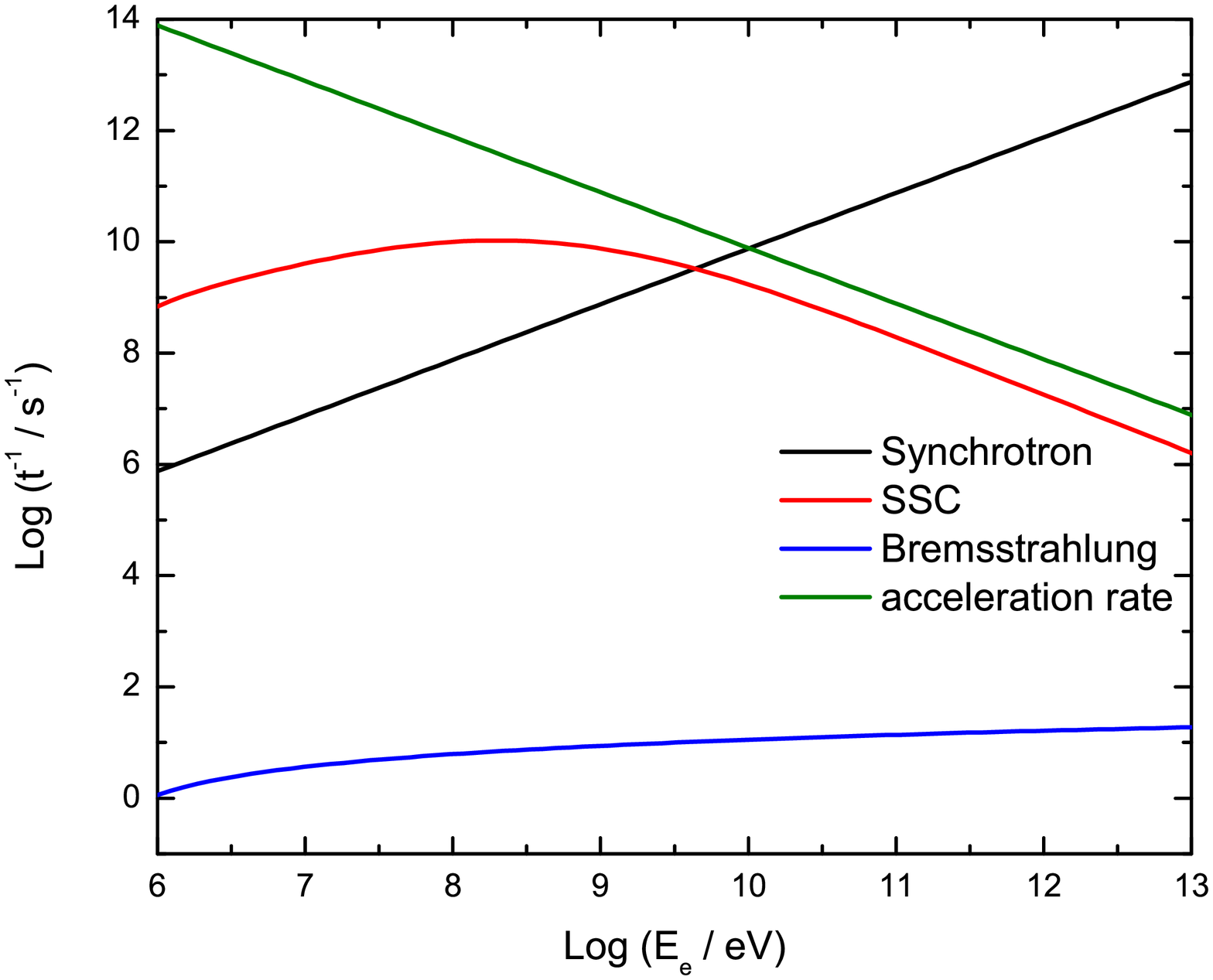,width=3in}}
 \hspace*{4pt}
 \parbox{3in}{\epsfig{figure=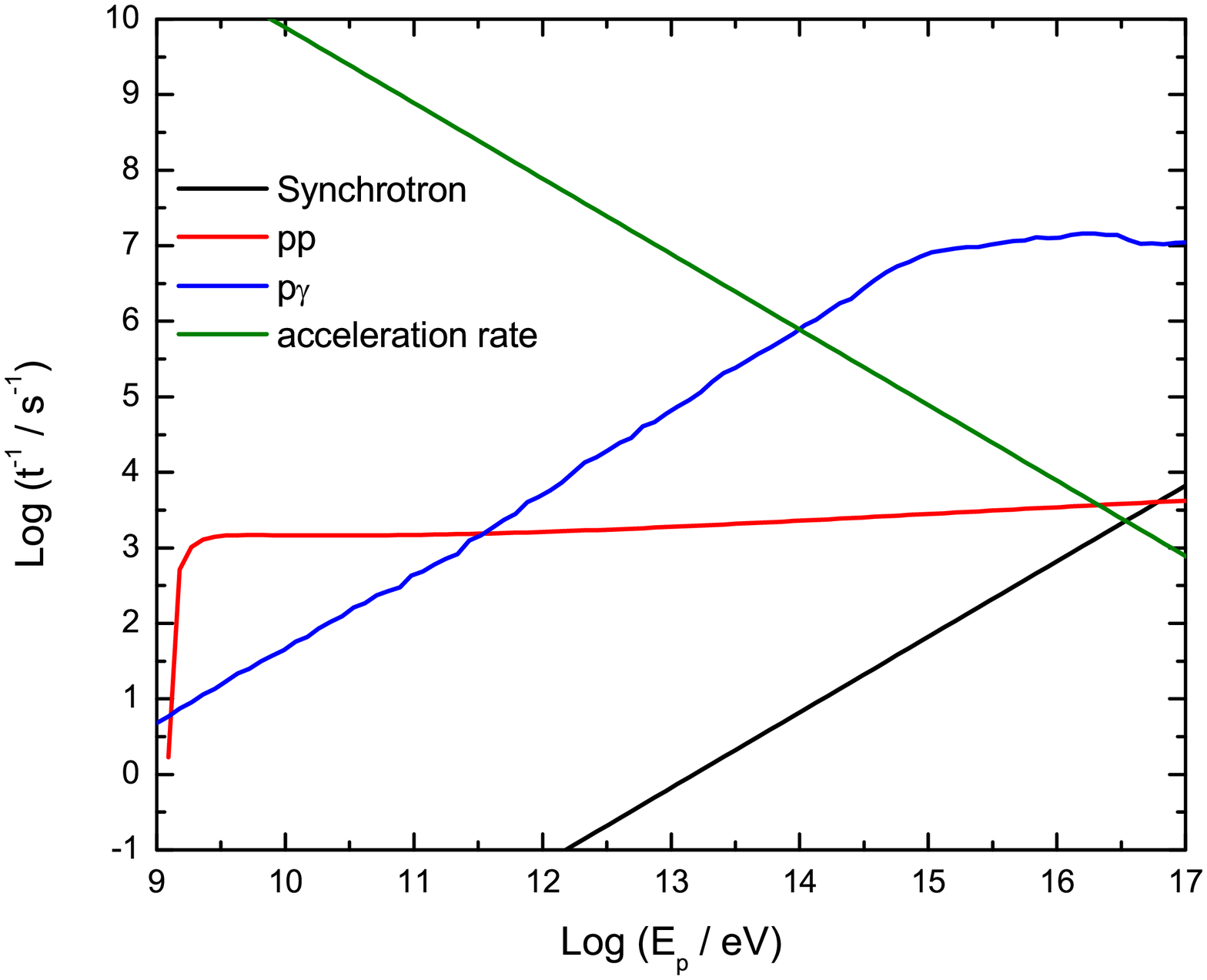,width=3in}}
 \caption{Energy losses and acceleration rate in the jet frame for electrons (left panel) and protons (right panel) in the reverse shock region, characterized by the parameters in Table \ref{table}.}
\label{fig:lossesRS}
\end{center}
\end{figure*}

Figure \ref{fig:lossesRS} shows electron and proton radiative losses, with the acceleration rate in the jet frame. The maximum energy for electrons and protons can be obtained by equating the acceleration and the cooling rates. The maximum energy that particles can attain in the RS region are $ 10^{10}$ eV and $10^{14}$ eV for electrons and protons, respectively. The synchrotron photons produced by electrons are the target for IC scattering (synchrotron self Compton, SSC), which is the main mechanism responsible for electron energy loss. The main mechanism of proton energy loss is photomeson production, and $pp$ interactions are relevant only for low- energy protons. 

The energies of neutrinos produced in the RS region can be estimated ad hoc. In this region, the main channel for energy loss of protons is photomeson production. There is a high peak in the photomeson production cross section at photon energies $\epsilon_{\rm{ph}} \sim 0.35$ MeV in the proton rest frame, owing to the $\Delta$-resonance \citep{stecker1973}. Most of the contribution to neutrino production then, comes through this channel. The condition that a proton must fulfill to create pions is \citep{zhangKumar2013}

\begin{equation}
E_{p}E_{\gamma} \sim 0.147 \rm{ GeV}^2 \Big( \frac{\Gamma}{1+z} \Big)^2.
\end{equation}

\noindent Neutrinos produced in $p\gamma$ interactions have energies of $E_{\nu} = 0.05 E_{p}$. In the reverse shock region, the target photon source is the synchrotron field produced by electrons, which has a peak at $E_{\gamma} \sim 100$ MeV. Then, photohadronic interactions would result in the production of TeV neutrinos.

\begin{figure*}[t]%
\begin{center}
 \parbox{3in}{\epsfig{figure=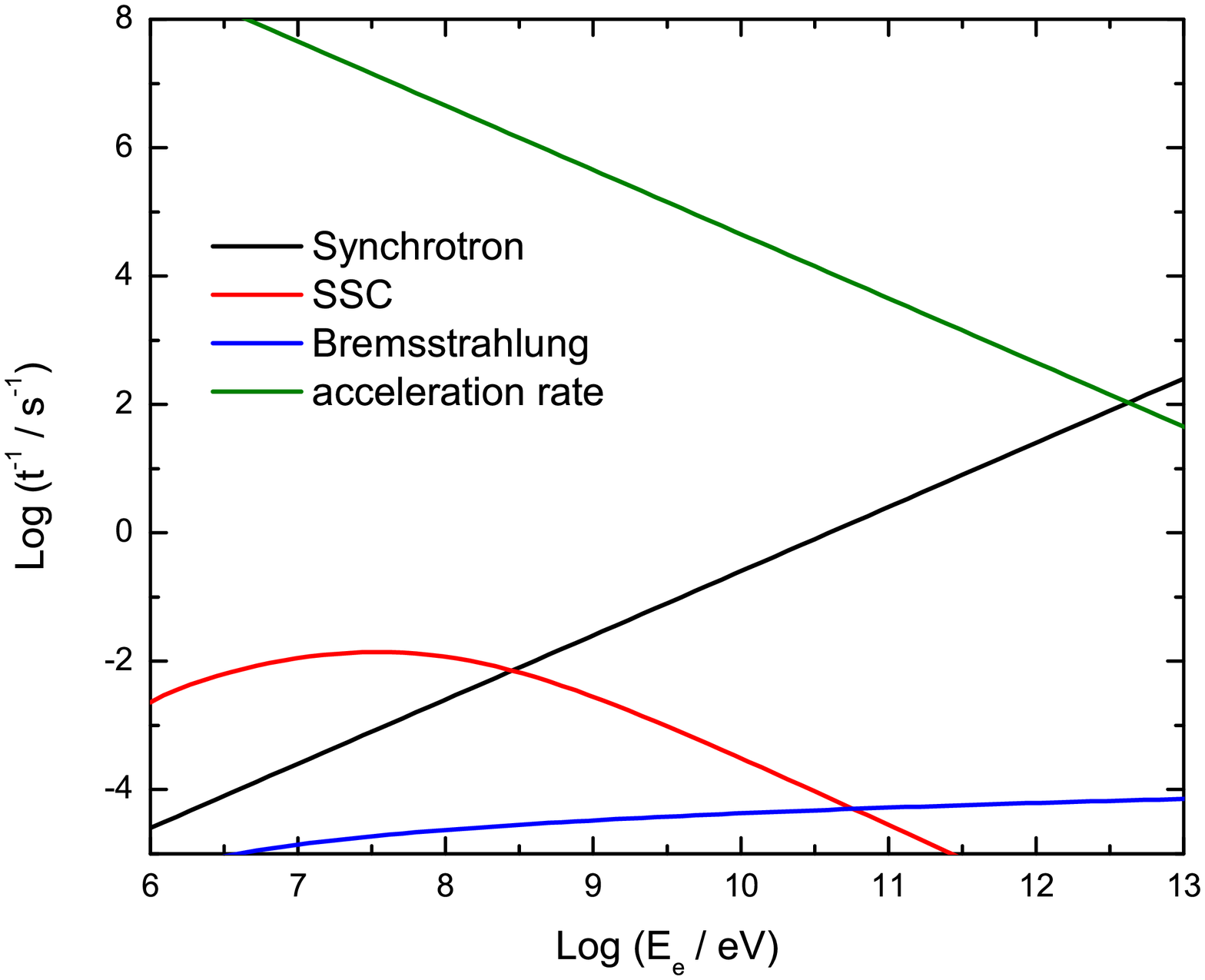,width=3in}}
 \hspace*{4pt}
 \parbox{3in}{\epsfig{figure=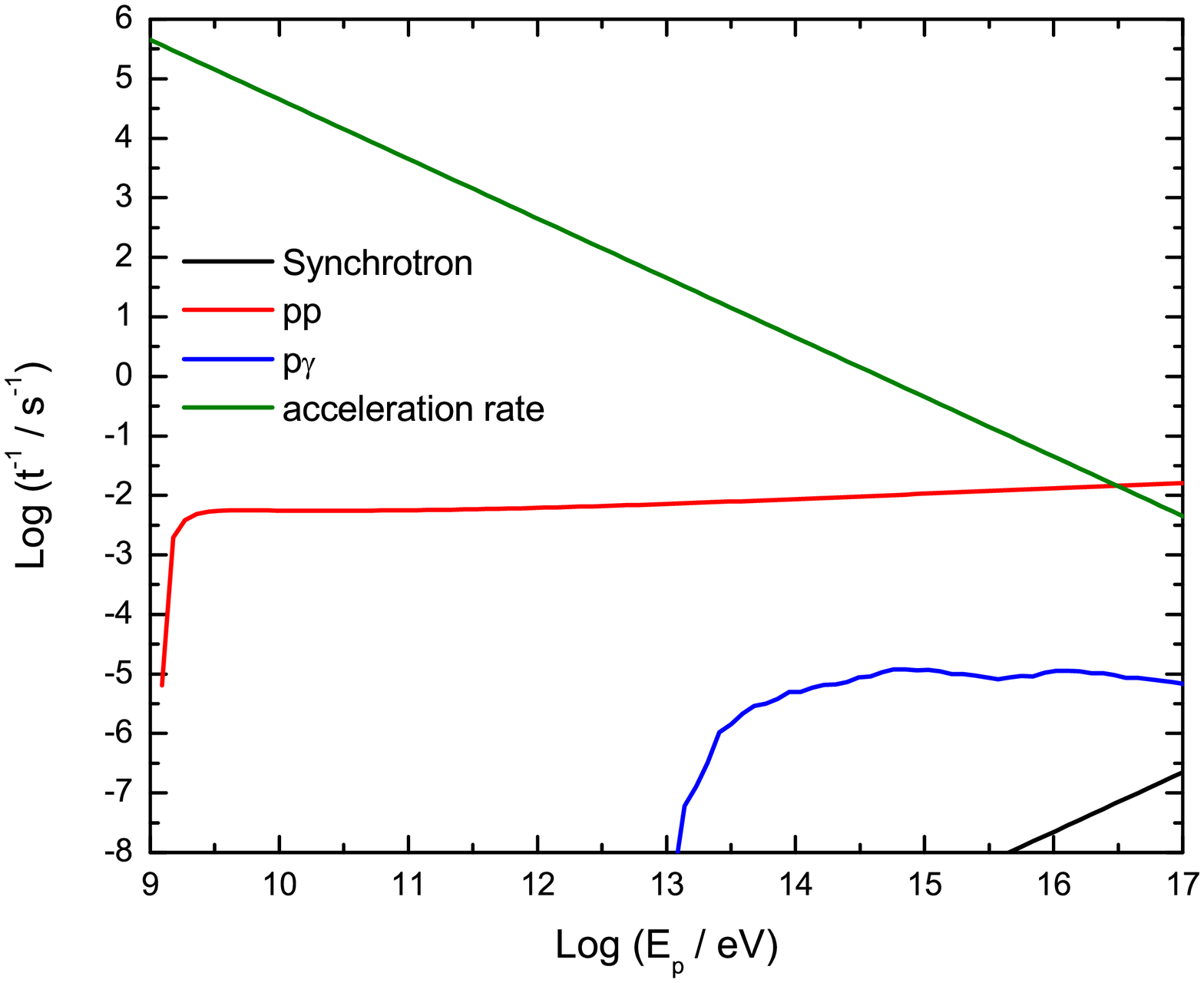,width=3in}}
 \caption{As in Fig. \ref{fig:lossesRS} but in the forward shock region.}
\label{fig:lossesFS}
\end{center}
\end{figure*}

Figure \ref{fig:lossesFS} shows the radiative losses in the FS region. Here, the main photon field is the stellar field. The mechanisms that dominate radiative losses are the same as in the RS region for electrons, whereas hadronic interactions play the main role for proton energy losses. The maximum energy achieved by electrons in this region is $4 \times 10^{12}$ eV, and the maximum energy of protons is determined not by radiative losses but by the size of the acceleration region \citep{hillas1984}, which results in $3 \times 10^{15}$ eV.

Since SSC and photohadronic interactions are the main channels for energy losses, transport equation for massive particles and for photons are coupled. The complete description of the method used for solving these equations can be found in \citet{vieyro2012}.

\subsection{Transport equations}

When protons and electrons are accelerated in both the FS and RS regions, they interact with the fields present in the source, thereby modifying their energy distributions and producing secondary particles. These secondary particles, moslty charged pions and muons, decay to produce neutrinos. To compute the neutrino injection as a result of pion and muon decay, it is necessary to first obtain the steady state particle distribution of these species. \citet{reynoso2009} have shown that radiative losses of the secondaries considerably affect the final neutrino fluxes in strongly magnetized sources. Then, to estimate the neutrino production in collapsars, we need to solve a set of differential transport equations, including those for transient particles.  

Because the synchrotron radiation produced by electrons provides a target photon field for photomeson production and IC scattering, the transport equations are coupled. In addition, the synchrotron radiation also absorbs the gamma rays emitted through hadronic interactions and, as a result, a second generation of electron-positron pairs is injected into the system. The steady state will be obtained on short timescales, due to the efficiency of the losses. 

The set of coupled transport equations is the following:

\begin{equation}\label{eq:electron}
                \frac{\partial }{\partial E} \left( b_{i}(E) N_{i}(E) \right)+ \frac{N_{i}(E)}{t_{\rm{esc}}} = Q_{i}(E),
        \end{equation} 
        
\noindent where $i=e^+$, $e^-$, $p$ refer to positrons, electrons, and protons, respectively, and

\begin{equation}\label{eq:pion}
                \frac{\partial }{\partial E} \left( b_{i}(E)  N_{i}(E) \right)+ \frac{N_{i}(E)}{t_{\rm{esc}}} + \frac{N_{i}(E)}{t^{i}_{\rm{dec}}} = Q_{i}(E),
        \end{equation} 

\noindent where $i=\pi^+$, $\pi^-$, $\mu^+$, $\mu^-$, refer to charged pions and muons, respectively. The last equation describes the transport of photons

\begin{equation}\label{eq:photon}
        \begin{aligned}
                \frac{N_{\gamma}(E_{\gamma})}{t_{\rm{esc}}} = Q_{\gamma}(E_{\gamma}) &+ Q_{e^{\pm} \rightarrow \gamma}(N_{e^{\pm}},E_{\gamma}) \\
                &- Q_{\gamma\gamma\rightarrow e^{\pm}}(N_{\gamma},E_{\gamma}) .
                \end{aligned}
        \end{equation} 

\noindent Here, $N_{i}(E)$ represents the steady state of each particle distribution (in units of erg$^{-1}$ cm$^{-3}$); $b_{i}(E)$ includes all radiative losses for a given type of particle; $t_{\rm{esc}}$ is the timescale over which relativistic particles escape from the system, which is taken as the crossing time; $t_{\rm{esc}} = r_{\rm{j}}(R_*) / c$, $t^{i}_{\rm{dec}}$ is the mean decay time for transient particles (pions and muons); and $Q_{i}(E)$ is the injection function. The term $Q_{\gamma}(E_{\gamma})$ represents photon injection by several radiative processes. The term $Q_{e^{\pm} \rightarrow \gamma}$ accounts for photons produced by pair annihilation, whereas $Q_{\gamma\gamma\rightarrow e^{\pm}}$ is a photon sink term due to photo-pair production. The expressions used for each process and the method used to solved the equations are detailed in \citet{vieyro2012}. The reader is referred to that paper for more details.

\section{Neutrino emission}\label{neutrinos}

Once the distributions of charged pions $N_{\pi}(E)$ and muons $N_{\mu}(E)$ are known, we proceed to estimate the neutrino emission of the source. We are interested in estimating the $\nu_{\mu}$ production, since the searches for point-like neutrino emission are optimized for this neutrino flavor. We also estimate the production of electron neutrinos in order to take the effects of neutrino oscillations into account. (The production of tau neutrinos is considered to be negligible initially.) Then, we consider $\nu_{e}$ production by the channel of muon decay,

        \begin{equation}
                \begin{aligned}
                 \mu^{-} & \rightarrow & e^{-} + \nu_{\mu} + \bar{\nu}_{e} ,\\
                 \mu^{+}& \rightarrow & e^{+} + \bar{\nu}_{\mu} + \nu_{e},
                \end{aligned}
                \end{equation}
                
\noindent and $\nu_{\mu}$ production by the previous channel plus charged pion decay:

                \begin{equation}
                \begin{aligned}
                 \pi^{-} &\rightarrow& \mu^{-} + \bar{\nu}_{\mu} , \\
                 \pi^{+}& \rightarrow &\mu^{+} + \nu_{\mu}.
                \end{aligned}
                \end{equation}
                
\noindent The current neutrino detectors cannot distinguish between neutrino and antineutrino, so we simply add both fluxes. Then, the total emissivity of muon neutrinos is

\begin{equation}
\begin{aligned}
\phi_{\nu_{\mu}+\bar{\nu}_{\mu}}(E)& =  \phi_{\pi^+ \rightarrow \nu_{\mu}}(E) + \phi_{\pi^- \rightarrow \bar{\nu}_{\mu}}(E) +\\
&+ \phi_{\mu^- \rightarrow \nu_{\mu}}(E) + \phi_{\mu^+ \rightarrow \bar{\nu}_{\mu}}(E).
\end{aligned}
\end{equation}

\noindent where

        \begin{equation}\label{eq:pi_mu}
        \begin{aligned}
                \phi_{\pi^+ \rightarrow \nu_{\mu}}(E,t) = & \int^{E^{\rm{max}}}_{E/(1-r_{\pi^+})}  dE_{\pi^+} \Big[ t^{-1}_{\pi^+,\rm{ dec}}(E_{\pi^+}) \\       
                 & \times N_{\pi^+}(E_{\pi^+},t) \frac{1 }{E_{\pi^+}(1-r_{\pi^+})} \Big],
                 \end{aligned}
        \end{equation}
        
\noindent with $r_{\pi}=(m_{\mu}/m_{\pi})^2$. The spectrum of $\bar{\nu}_{\mu}$ produced by the decay of $\pi^-$  is also described by Eq. \ref{eq:pi_mu}.

For the decay of muons,

\begin{equation}\label{eq:mu_nu}
        \begin{aligned}
                \phi_{\mu^- \rightarrow \nu_{\mu}}(E,t) &= \sum^2_{i=1} \int^{E^{\rm{max}}}_{E} dE_{\mu} t^{-1}_{\mu,\rm{ dec}}(E_{\mu})N_{\mu_{i}}(E_{\mu},t)\\
                 &\times 
                 \frac{ F_{\mu \rightarrow \nu_{\mu}} (E / E_{\mu} )}{E_{\mu}},
                \end{aligned}
        \end{equation}
        
\noindent where
        
        \begin{equation}
        \begin{aligned}
                F_{\mu \rightarrow \nu_{\mu}} (x) =      & \left(\frac{5}{3}-3x^2+ \frac{4}{3}x^3\right) + \\
                 & + h\left(-\frac{1}{3}+3x^2 - \frac{8}{3}x^3\right) . 
                 \end{aligned}
                \end{equation}
                
\noindent In this expression, $x=E/E_{\mu}$, $\mu_{\{1,2\}}=\mu^-_{\rm{L,R}}$. The production of $\bar{\nu}_{\mu}$ by the decay of $\mu^+$ is similar but replaces $\mu_{\{1,2\}}=\mu^+_{\rm{L,R}}$. The values of the helicities are $h_{(\mu^-_{\rm{L}},\mu^+_{\rm{L}})} = - h_{(\mu^-_{\rm{R}},\mu^+_{\rm{R}})} = -1$.

In a similar way to Eq. \ref{eq:mu_nu}, the total emissivity of electron neutrinos $\nu_{e}$ is \citep{lipari2007}

        \begin{equation}
        \begin{aligned}
                \phi_{\mu^+ \rightarrow \nu_{e}}(E,t) = \sum^2_{i=1}  \int^{E^{\rm{max}}}_{E} & dE_{\mu} t^{-1}_{\mu,\rm{dec}}(E_{\mu})N_{\mu}(E_{\mu},t) \\
                &\times \frac{ F_{\mu \rightarrow \nu_{e}} (E / E_{\mu} )}{E_{\mu}},
                \end{aligned}
        \end{equation}
        
\noindent where
        
        \begin{equation}
        \begin{aligned}
                F_{\mu \rightarrow \nu_{e}} (x) =  &\left(2-6x^2+ 4x^3\right) + \\
                & + h\left(2-12x+18x^2 - 8x^3\right)  .
        \end{aligned}
        \end{equation}

\subsection{Spin precession}

Within the context of the Standard Model, neutrino are massless particles, without electric charge and thus have no magnetic moment. Then, in principle, they do not interact electromagnetically. However, in a minimal extension of the Standard Model in which neutrinos become massive \citep{mohapatra1991}, the electroweak coupling between neutrinos and $W$ bosons has the same effect as an effective electric charge (neutrinos can interact with a photon through radiative loop diagrams), which induces a magnetic moment \citep{mohapatra1991}. The effect of a nonzero magnetic moment is to rotate the spin of the neutrino in the presence of a magnetic field, that is, to change the helicity of the neutrino. This effect is known as neutrino spin-flavor precession (SFP), and was proposed by \citet{akhmedov2002} as a secondary mechanism responsible for the deficit of $\nu_e$ solar neutrinos (the main mechanism is standard neutrino oscillation).

This mechanism has been recently proposed as a possible explanation of the negative results in the search for ultra-high energy neutrinos. In the case of solar neutrinos, this process has been shown not to be very efficient; in AGNs and GRBs, however, given the range of magnetic field intensities and sizes of the sources, for a reasonable value of the neutrino magnetic moment, a spin transition could be induce \citep{barranco2012}.

The magnetic moment of a Dirac neutrino in the Standard Model of particle physics is $\mu_{\rm{Q}} = eG_{\rm{F}}m_{\nu} \sim 3.2 \times 10^{-19} (m_{\nu}/ 1 {\rm eV})\mu_{\rm{B}}$, where $G_{\rm{F}}$ is the Fermi constant and $m_{\nu}$ is the neutrino mass. Contrary to the standard magnetic moment of charged particles, which is inversely proportional to the mass, this induced magnetic moment depends linearly on the neutrino mass. 

We consider two cases of transitions: one is the conversion due to a diagonal magnetic moment that changes the active electron neutrino into a righthanded sterile electron neutrino, also called horizontal transition,

\begin{equation}
\nu_{e_{\rm{L}}} \rightarrow  \bar{\nu}_{e_{\rm{R}}} ,
\end{equation}

\noindent and the other case is due to the non-diagonal magnetic moment called vertical transition. It acts between different flavors of neutrinos:

\begin{equation}
\bar{\nu}_e  \leftrightarrow   \nu_{\mu} .
\end{equation}

Then, taking the probability of spin transition into account, the neutrino flux emerging from the source is

\begin{equation}
\phi_{\nu_{e}}  = P(\nu_{e_{\rm{L}}} \rightarrow \nu_{e_{\rm{L}}}) \phi^0_{\nu_{e}} + P(\bar{\nu}_{\mu} \rightarrow \nu_{e}) \phi^0_{\bar{\nu}_{\mu}} ,
\end{equation}

\begin{equation}
\phi_{\nu_{\mu}}  = P(\nu_{\mu_{\rm{L}}} \rightarrow \nu_{\mu_{\rm{L}}}) \phi^0_{\nu_{\mu}} + P(\bar{\nu}_{e} \rightarrow \nu_{\mu}) \phi^0_{\bar{\nu}_{e}} ,
\end{equation}

\noindent where the conversion probabilities are given by

\begin{equation}
\begin{aligned}
P(\bar{\nu}_{\mu} \rightarrow \nu_e, r ) = & P(\bar{\nu}_e \rightarrow \nu_{\mu}, r ) = \\
=&  \sin^2 \Big( \int_0^r \frac{\mu_{\nu} B_{\perp} (r')}{ \hbar c} dr' \Big),
\end{aligned}
\end{equation}

\noindent and 

\begin{equation}
 P(\nu_{e_{\rm{L}}} \rightarrow \nu_{e_{\rm{L}}}, r ) = 1 - P(\bar{\nu}_{\mu} \rightarrow \nu_e, r ).
\end{equation}

\noindent Similar equations apply to $\phi_{\bar{\nu}_{\mu}}$ and $\phi_{\bar{\nu}_{e}}$.

The current limits to the neutrino magnetic moment are $\mu_{\nu} \leq 10^{-11} \mu_{\rm{B}}$,  coming from laboratory measurement or from a combined analysis, and $\mu_{\nu} \leq 10^{-12} \mu_{\rm{B}}$, from astrophysical observations or from solar data \citep{akhmedov2002}. Here we adopt $\mu_{\nu} \leq 10^{-12} \mu_{\rm{B}}$.

For our collapsar model, in the RS region, the probability can be approximated as

\begin{equation}
P(\bar{\nu}_{\mu} \rightarrow \nu_e)  =  \sin^2 \Big(  \frac{\mu_{\nu} B_{\perp} (R_{\star}) R_{\rm{jet}}(R_{\star}) }{ \hbar c}  \Big) \approx 0.78 ,
\end{equation}

\noindent therefore,

\begin{equation}
 P(\nu_{\rm{L}} \rightarrow \nu_{\rm{R}}) \approx 0.22 .
\end{equation}

\subsection{Standard neutrino oscillations}

It is well known that neutrinos can oscillate between three distinct flavors: muon, electron, and tau neutrino. This can affect the final flux of neutrinos of a given flavor. For astrophysical sources at very high distances, the arriving flux on Earth is \citep{esmaili2010}

\begin{equation}
\phi_{\alpha} = \sum_{\beta = e, \mu, \tau} P_{\alpha \beta} \phi^{0}_{\beta},
\end{equation}

\noindent where $\phi^{0}_{\alpha}$ is the neutrino flux of flavor $\alpha$ at the source, and $P_{\alpha \beta}$ is the oscillation probability. This is given by

\begin{equation}
P_{\alpha \beta} = \sum_{j=1}^{3} |U_{\alpha j}|^2|U_{\beta j}|^2 .
\end{equation}

\noindent Here $U_{\alpha j}$ is the mixing matrix. The values of the mixing matrix depend on the standard oscillation parameters, the mixing angles $\theta_{12}$, $\theta_{23}$, and $\theta_{13}$. The current best fit values ($3\sigma$) for these parameters are \citep{esmaili2010}

\begin{equation}
        \begin{aligned} 
                \sin ^2 \theta_{12} &=& 0.27 - 0.34 , \\
                \sin ^2 \theta_{23} &=& 0.34 - 0.67 , \\
                \sin ^2 \theta_{13} &=& 0.016 - 0.030.
        \end{aligned}
\end{equation}

The final values of the mixing matrix are taken from \citet{vissani2011}, and the final neutrino flux results in

\begin{eqnarray}
\phi_{\nu_{\mu}} & = & P_{\mu e} \phi^0_{e} + P_{\mu \mu} \phi^0_{\mu} + P_{\mu \tau} \phi^0_{\tau} \\
           & = & 0.221 \phi^0_{e} + 0.390 \phi^0_{\mu} + 0.390 \phi^0_{\tau}.
\end{eqnarray}

\subsection{Neutrino fluxes on Earth}\label{detectability}

\begin{figure*}
\begin{center}
 \parbox{3in}{\epsfig{figure=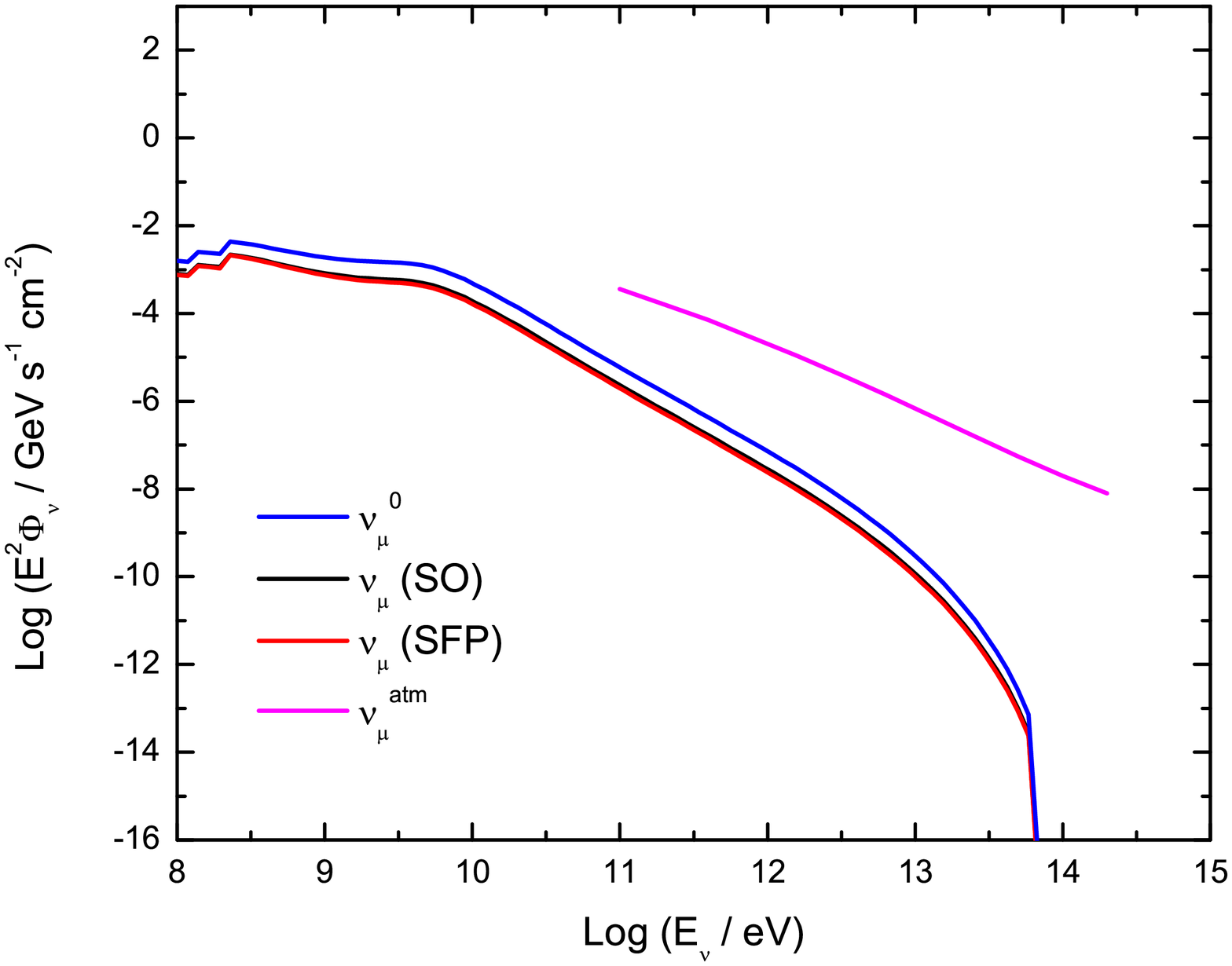,width=3in}}
 \hspace*{4pt}
 \parbox{3in}{\epsfig{figure=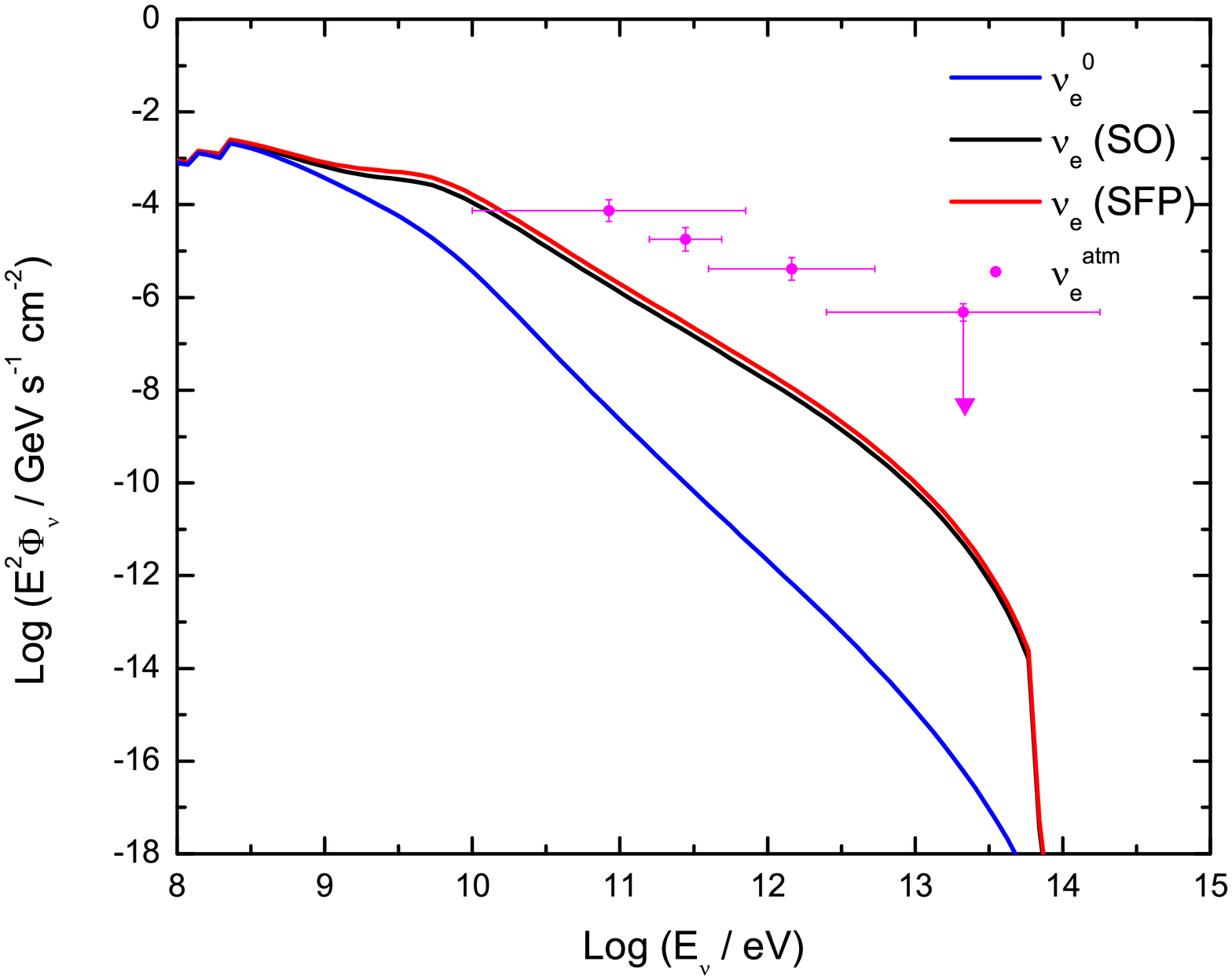,width=3in}}
 \caption{Flux of neutrinos arriving at Earth (blue line) produced in the reverse shock region. The black line is the final neutrino flux after standard oscillations (SO), whereas the red line is the flux taking the coupling between the neutrino magnetic momentum and the magnetic field (SFP) into account.}
\label{fig:neutrinosRS}
\end{center}
\end{figure*}

\begin{figure*}
\begin{center}
 \parbox{3in}{\epsfig{figure=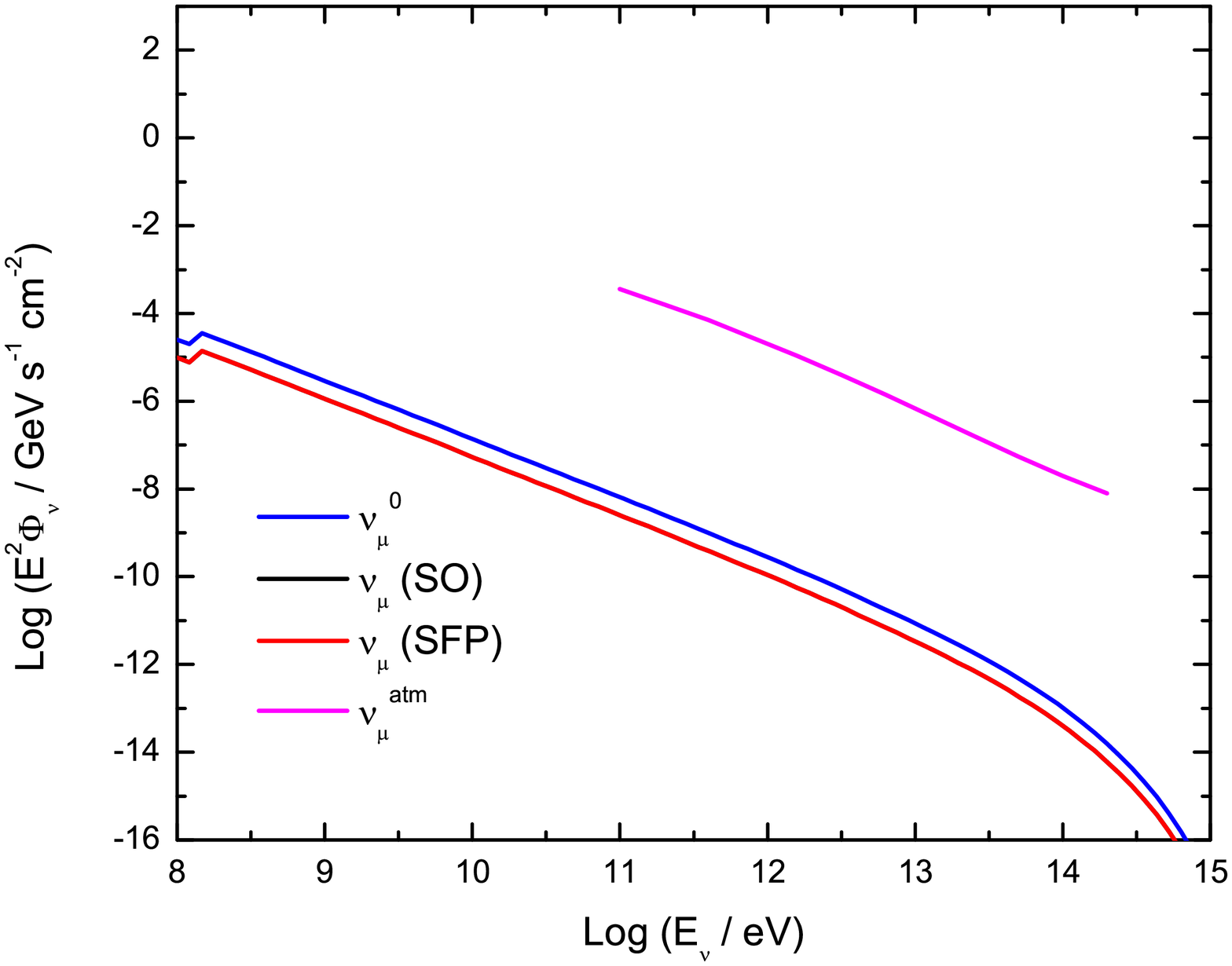,width=3in}}
 \hspace*{4pt}
 \parbox{3in}{\epsfig{figure=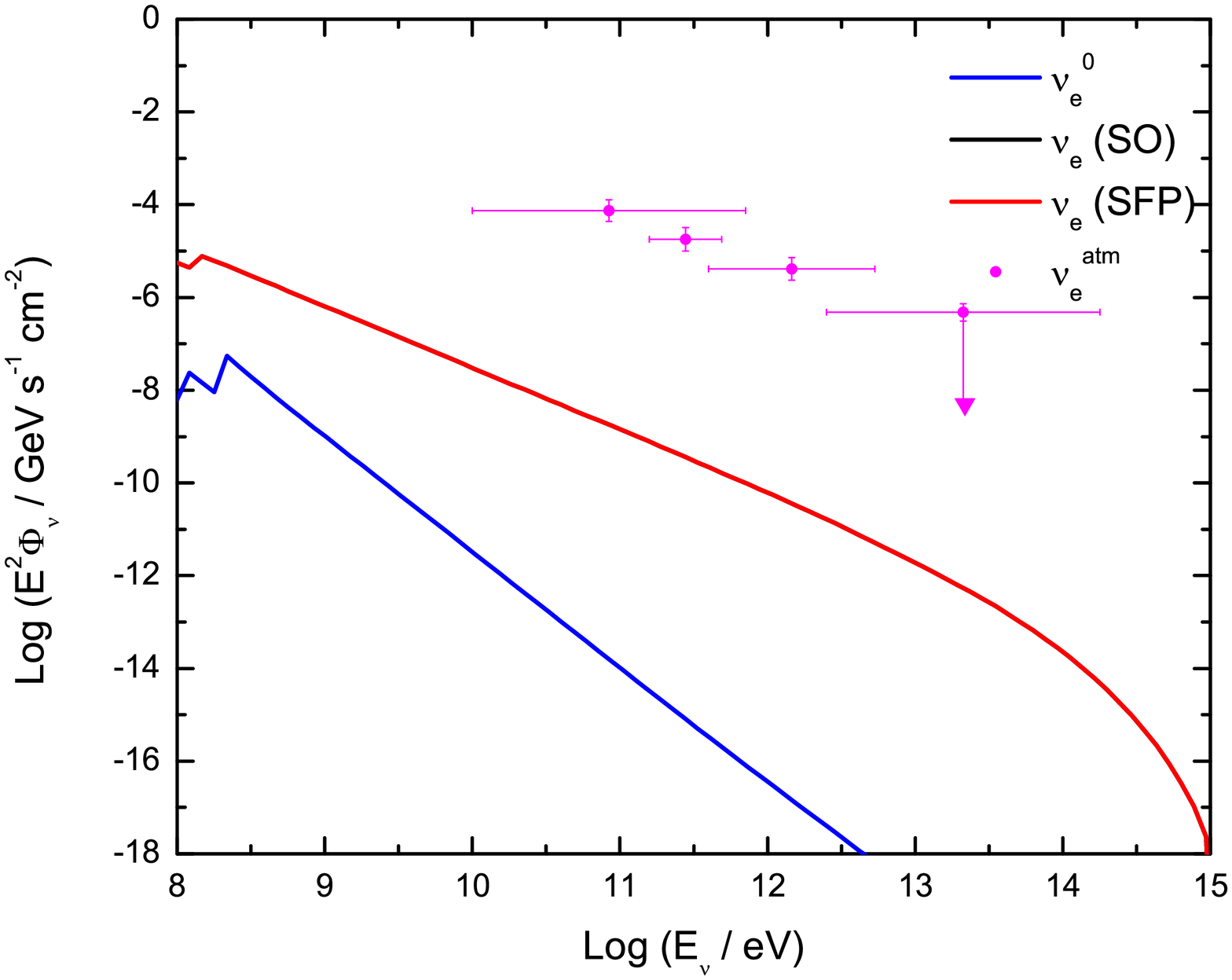,width=3in}}
 \caption{As in Fig. \ref{fig:neutrinosRS}, but for the forward shock region.}
\label{fig:neutrinosFS}
\end{center}
\end{figure*}

The differential flux of neutrinos arriving at the Earth can be obtained as

\begin{equation}
\frac{d \Phi_{\nu_{\mu}}}{dE} = \frac{D}{4\pi d^2} \int_{\rm{V}}{d^3r \phi_{\mu}(E,t)} ,
\end{equation}

\noindent where $D^{-1} = \Gamma(1-\beta)$ \citep{reynoso2012}. We consider a nearby event at $z \sim 0.2$. Figures \ref{fig:neutrinosRS} and \ref{fig:neutrinosFS} show this quantity, weighted by the squared energy in the RS and FS regions, respectively. The figures also show the neutrino flux affected by standard oscillations (SO) and spin flavor precession (SFP).

The changes in the neutrino flux from the SO case can be significant. The effects of the neutrino SFP, however, are not relevant in this scenario. In the FS, where the value of the magnetic field is low, this effect is completely negligible (see Fig. \ref{fig:neutrinosFS}).

\subsection{Event rate and detection with IceCube}\label{eventRate}

The number of events detected by IceCube is

\begin{equation}
N = \int_{z} \int_{E_{\nu}} \Phi_{\nu} A_{\rm{eff}} (E_{\nu},z) dE_{\nu} d\cos(z).
\end{equation}

\noindent We adopt the values for the IceCube effective area from \citet{IceCubeAtmo2011}; we use the effective area in ten bins of $\cos (z)$ (with bin width $0.1$) and 12 bins of $E_{\nu}$ (with bin width 0.3 in $\log(E_{\nu}/GeV)$, \citealt{esmaili2012}). 

The number of events of atmospheric muon neutrinos ($\nu_{\mu} + \bar{\nu}_{\mu}$), from 100 GeV to 400 TeV, detected by IceCube is $\sim 17700$ events per year \citep{IceCubeAtmo2011}. The atmospheric muon  and electron neutrino energy spectrum are also shown in Figs. \ref{fig:neutrinosRS} and \ref{fig:neutrinosFS} \citep{IceCubeAtmo2011,IceCubeElecAtmo2012}. That is equivalent to an event rate of $5.6 \times 10^{-4}$ Hz.

For the RS region, above 2 GeV, the total neutrino event rate is $2.4 \times 10^{-7}$ Hz considering only SOs, and $2.08 \times 10^{-7}$ Hz taking into account SPF. For the FS region, the total neutrino event rate is $6 \times 10^{-10}$ Hz for both cases (SOs only and with SFP effects). These values are considerably lower than those obtained in the RS. This was expected, since in the FS region photohadronic interactions are not important.

After integrating over one year of observations, the neutrino event rate for the RS region would be $\sim 7-8$. Taking into account that the rate of GRB events with $z < 0.5$ that can be detected from Earth is estimated to be $\sim 10^{3}$ per year \citep{meszarosWaxman2001}, then the number of muon events in one year would be  $\sim 7\times 10^{3}$, which is comparable to the atmospheric muon events detected by IceCube \citep{IceCubeAtmo2011}. A multi-year integration can then provide a detectable flux.

\section{Application to GRB progenitors from Pop. III}\label{popIII}

Recent works by \citet{meszarosRees2010}, \citet{gao2011}, and \citet{berezinsky2012} have extended the calculations of neutrino emission to Pop. III GRBs. These events are of particular cosmological interest, since they are related to the first stars formed in the universe. These stars are supposed to have been very massive, and accretion onto very massive black holes might lead to a scaled-up collapsar gamma-ray burst \citep{meszarosRees2010}. Here, we apply the same model as presented in the previous sections to GRBs with progenitor stars from Pop. III.

There are two types of Pop. III stars: the first group are the Pop. III.1 stars, which are formed by purely cosmological initial conditions; the second group, called Pop. III.2 stars, are assumed to be formed from zero-metallicity gas in the pre-ionization era. Recent numerical studies of Pop. III.2 stars show that these are not as massive as was once thought; the typical final values for the masses are $40-60 M_{\odot}$, because the original hydrogen clouds are very prone to fragmentation  \citep{smith2010,greif2011}.

On the other hand, Pop. III.1 stars are formed at $z>20$ and are supposed to have masses of $60-320 M_{\odot}$ \citep{norman2010}, except for those in the range $140-260 M_{\odot}$ which are subject to pair instability. The Pop. III.1 stars are expected to undergo a core collapse leading directly to a central black hole \citep{heger2002}, whose mass would be several tenths of a solar mass.

For these massive stars, we assume an efficiency of $q_{\rm{j}} = 0.2$. In this case, the luminosity results in $L_{\rm{j}} \sim 1.8 \times 10^{52}$ erg s$^{-1}$, in agreement with the values obtained by \citet{meszarosRees2010} for Pop. III GRBs (scaled for a $M_{\rm{BH}} = 10 M_{\odot}$). We consider a star radius of $R_{*} = 10^{13}$ cm; we then scaled the parameters that depend on these quantities (e.g., magnetic field at the surface of the star, power injected in relativistic particles, etc). The remaining parameters are the same as in Table \ref{table}. The radiative losses are similar to the case of WR stars, so we do not include these plots here. We only point out that particles can achieve slightly higher energies in this context, such as those in WR-GRBs.

\begin{figure*}
\begin{center}
 \parbox{3in}{\epsfig{figure=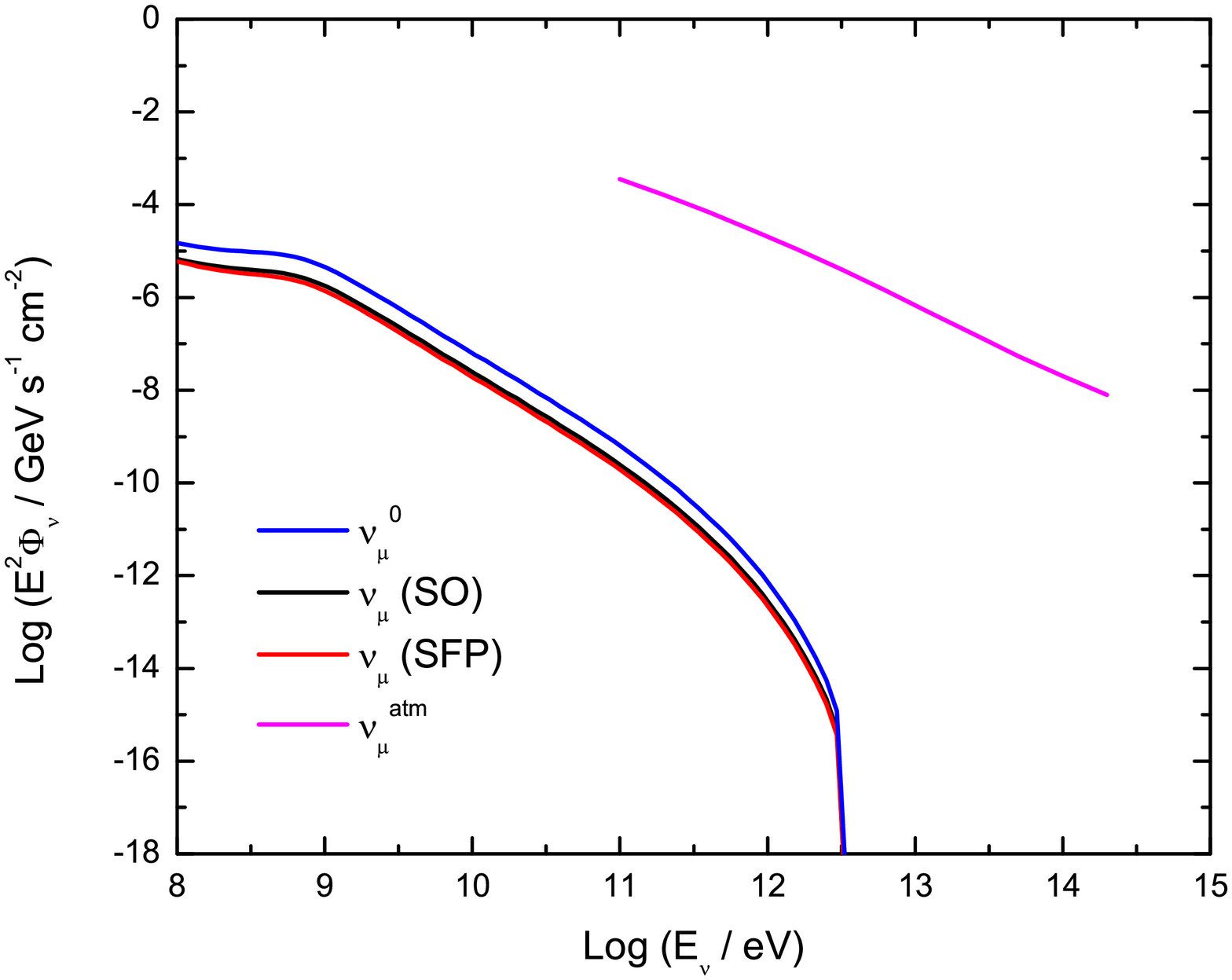,width=3in}}
 \hspace*{4pt}
 \parbox{3in}{\epsfig{figure=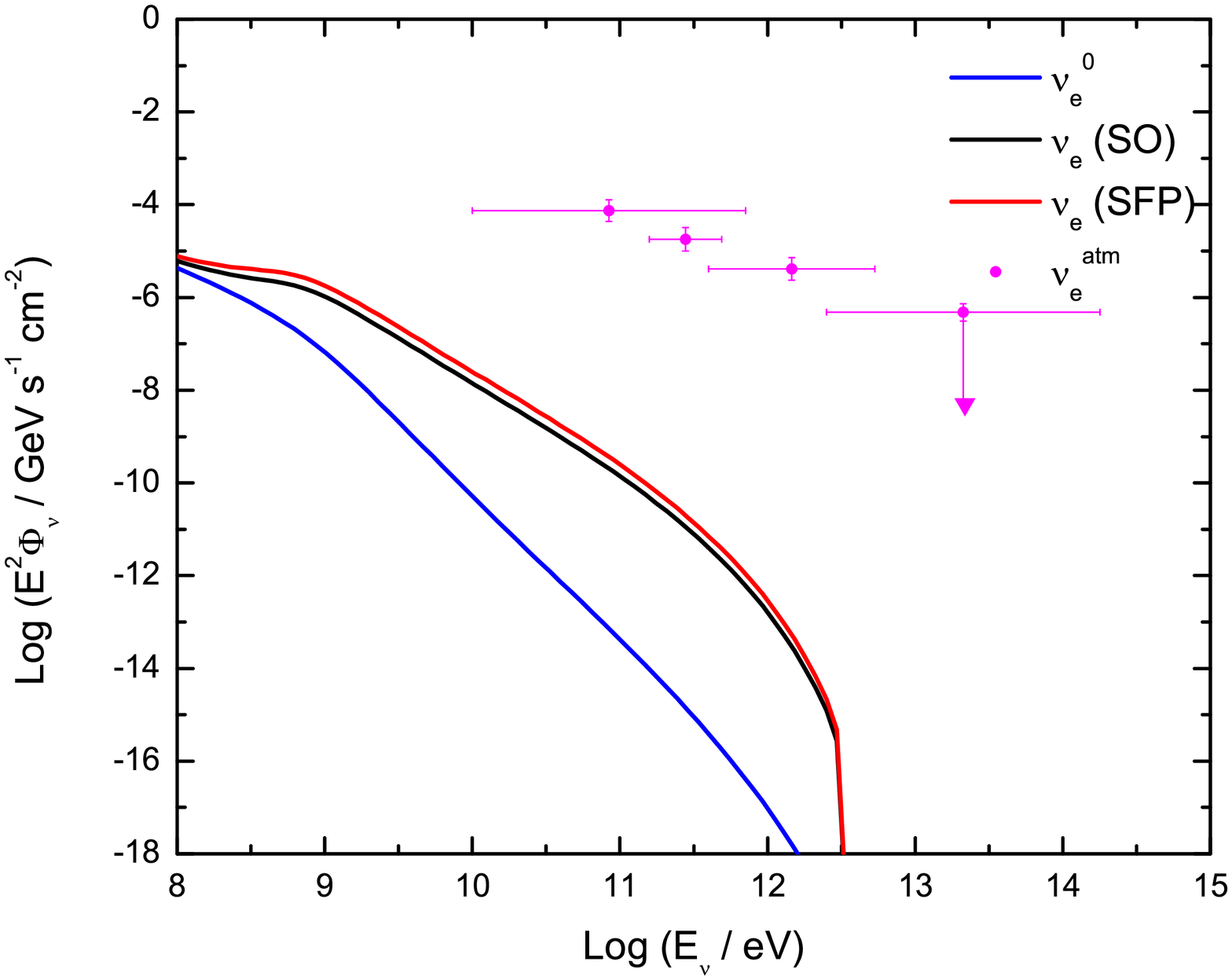,width=3in}}
 \caption{Flux of muon neutrinos (left panel) and electron neutrinos(right panel) arriving at Earth, produced in the reverse shock region, for GRBs with progenitors from Pop. III stars. The colors of the lines are the same as in Fig. \ref{fig:neutrinosRS}.}
\label{fig:neutrinos_popIII_RS}
\end{center}
\end{figure*}

\begin{figure*}
\begin{center}
 \parbox{3in}{\epsfig{figure=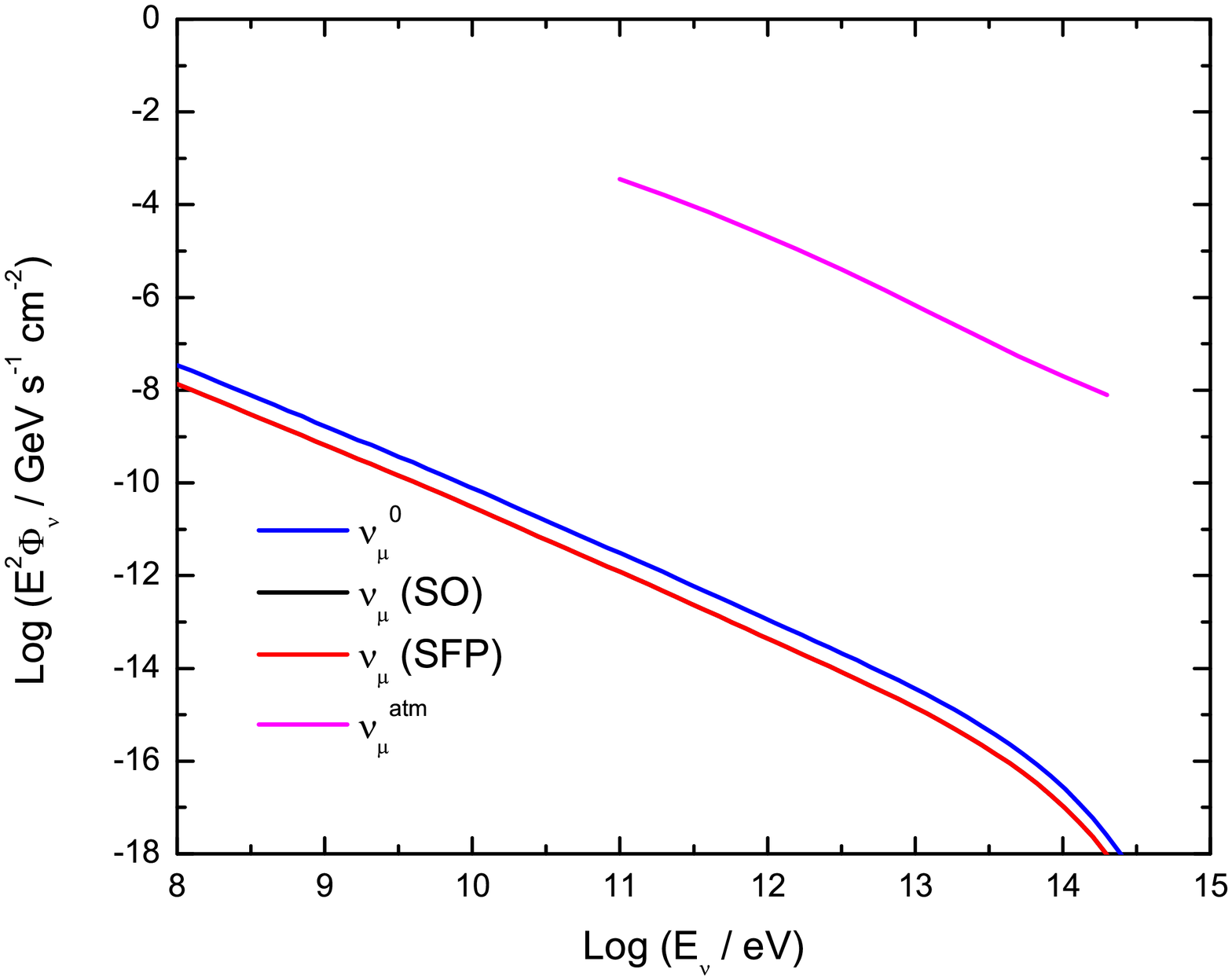,width=3in}}
 \hspace*{4pt}
 \parbox{3in}{\epsfig{figure=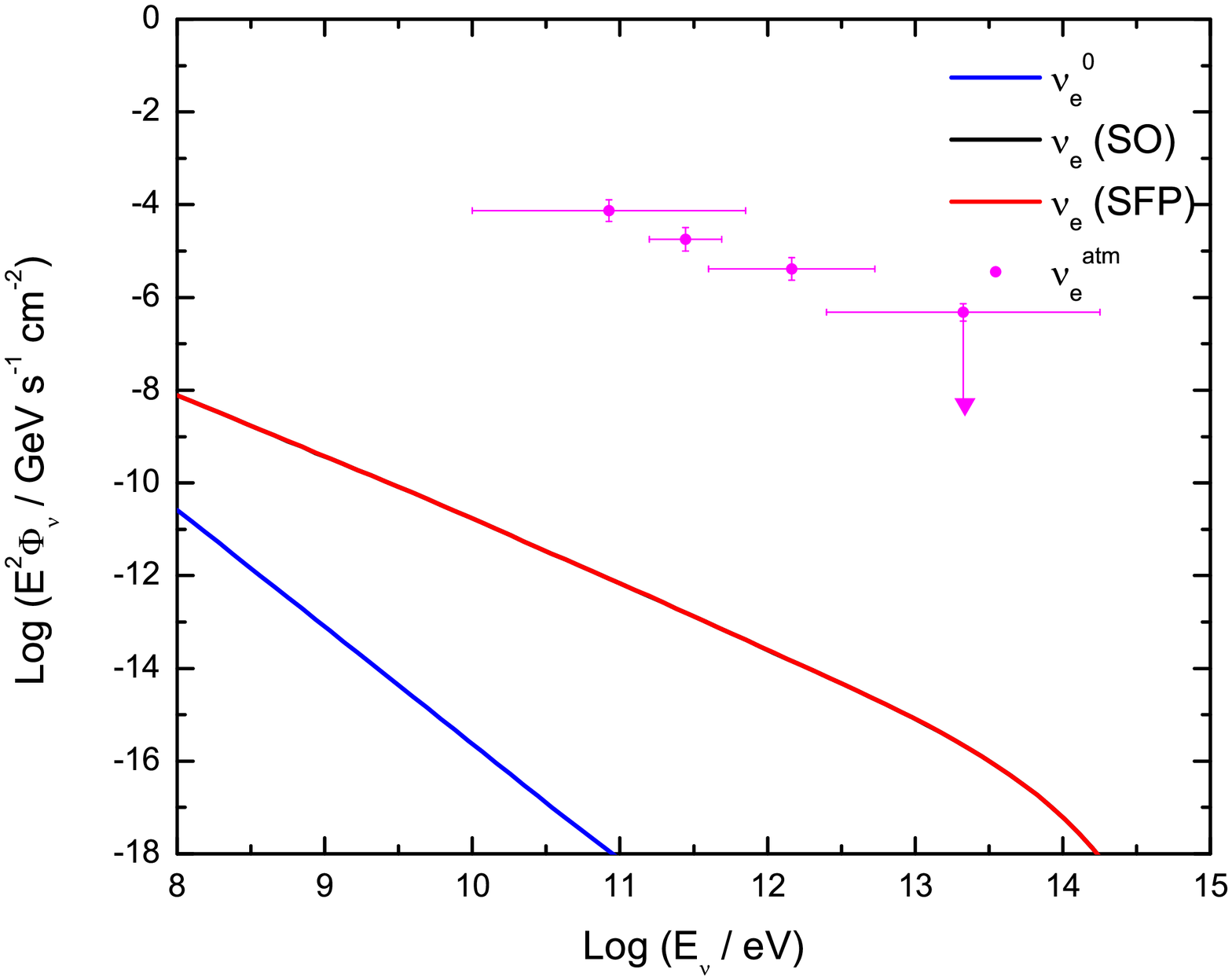,width=3in}}
 \caption{As in Fig. \ref{fig:neutrinos_popIII_RS}, but for the forward shock region.}
\label{fig:neutrinos_popIII_FS}
\end{center}
\end{figure*}

In Fig. \ref{fig:neutrinos_popIII_RS} we show the fluxes of muon (left panel) and electron neutrinos (right panel) arriving at Earth, produced in the RS region and obtained for GRBs with progenitors from Pop. III stars. We consider a GRB at $z=20$. The neutrino event rate is $1.4 \times 10^{-11}$ Hz and $1.1 \times 10^{-11}$ Hz, considering standard oscillations only and including spin precession effects, respectively. Figure \ref{fig:neutrinos_popIII_FS} shows the fluxes of muon neutrinos (left panel) and electron neutrinos (right panel) arriving at Earth, produced in the FS region. The neutrino event rate for this region is $2.2 \times 10^{-13}$ Hz for both SOs alone and with SFP effects.

According to our model, the neutrino emission from these shocks of a single Pop. III GRB would not be detectable by IceCube. The main reason is the great distance at which these events take place. The number of GRBs per year expected to be observed is  $N < 20$ GRBs, integrated over at $z > 6$ for Pop. III.2 and $N < 0.08$ per year integrated over at $z > 10$ for Pop. III.1 \citep{souza2011}. The model presented in this work can be applied to both Pop. III.1 and Pop. III.2 stars. Although Pop. III.1 are more powerful, given the higher redshift and mostly the expected number of events, Pop. III.2 are more likely to be detected. However, given that the maximum energy of particles is a few TeV, the background of atmospheric neutrinos is the dominant component in the operational range of IceCube.

\section{Discussion}

There are a few factors responsible for the low neutrino event rates we have obtained in our work. The radiative losses in our model are not negligible and the maximum energies that particles, in particular protons, can achieve do not exceed a few TeV. This is a consequence of considering the effects of the high magnetic fields that there should be in the sources. The neutrino flux in the energy range where IceCube is effective decreases considerably in comparison to previous estimates, causing the event rate to be low. It has already been pointed out by \citet{li2011} that some theoretical predictions (see e.g., \citealt{waxman1997,gao2011}) have overestimated the neutrino fluxes. This is mainly because these works do not consider the cooling of secondary particles such as pions and muons. Another simplification in the neutrino flux estimates is to ignore the energy dependence of charged pion and muon production (e.g., \citealt{berezinsky2012}), which has a direct effect on the neutrino energy distribution.

Another relevant factor is the redshift. We first considered GRBs with $z \sim 0.2$. When we applied the same model for GRBs with progenitors from Pop. III at $z \sim 20$, the neutrino event rate decreased considerably.


One of the free parameters in our model is the ratio between the lepton and hadron content in the jet. Here we have adopted $a=0.1$. The neutrino flux scales approximately linear with this parameter. If the hadron content of the jet is smaller, then the event rate of neutrino will also decrease in approximately the same order of magnitude. Heavier jets are unlikely to be accelerated to high Lorentz factors. 

The evolution of the magnetic field along the jet is represented by Eq. \ref{eq:magneticField}, with an index $m$ that can vary between $1$ and $2$. The results shown in this paper were obtained assuming $m=1$. For $m > 1,$ the magnetic field in the surface of the star is lower, and this causes the acceleration rate to be lower, and the relativistic particles are not able to accelerate efficiently. Then, for higher values of $m$ the neutrino event rate decreases. It is important to notice that we use Eq. \ref{eq:magneticField} as a prescription for the evolution of the magnetic field in the jet; a more realistic characterization requires a detailed description of the model involved for the prompt emission of the GRB.

In this work we aim at estimating neutrino emission, regardless the mechanism responsible for the prompt gamma-ray emission (e.g., internal shocks, photospheric dissipation, etc). However, the details of the microphysics would inevitably affect the results presented here. For a complete discussion on how the different models may change the neutrino signals can be found in \citet{zhangKumar2013}.

\section{Conclusions}

We have investigated the neutrino production in the FS and RS regions at the surface of collapsars related to Wolf-Rayet and Pop. III stars. Given the efficiency of the radiative losses, we considered both regions to be in steady state during the duration of the event. We solved the set of coupled transport equations to determine the final particle distributions, and finally we estimated the neutrino emission for each model. 

We focused our study on the uncorking region close to the stellar surface. We cannot rule out possible interactions between the jet and inhomogeneities in the stellar medium as an additional neutrino source. For example, it has been proposed by \citet{barkana2000} that the first stars formed in the universe ionized the intergalactic medium, producing HII regions around them. Then, a possible scenario would be the jet interacting with the shell that surrounds the HII region, where the neutrino production may be enhanced by a high- density external material. The complexities of jet-cloud interactions have recently been explored by \citet{araudo2010}, at low Lorentz factors. 

We also studied how some effects in the context of reasonable extensions of the Standard Model can affect the intrinsic neutrino flux produced in these sources. In particular, we have seen that standard neutrino oscillations can play an important role in changing the different neutrino flavor fluxes, whereas SFP is almost negligible in all cases studied here.

We have found that the inclusion of radiative losses for particles and a self-consistent treatment for the transport of particles and photons
significantly reduce the neutrino event rate expected from these sources, which has been overestimated in previous works. Our results are in accordance with the non-detection of high-energy neutrino from GRBs by IceCube so far. However, our model for neutrino production in the reverse shock region of GRBs with $z <1$ suggests that we may be close to measuring the cumulative effect of these extragalactic sources. We conclude that the detection of neutrinos from collapsars requires long (timescales of $\sim$ years) integrations with the full IceCube array to surpass the atmospheric background.

\section*{Acknowledgments}

This work was partially supported by the Argentine agencies ANPCyT (BID 1728/OC -AR PICT-2012-00878) and CONICET (PIP 0078/2010), as well as the Spanish grant AYA 2010-21782-C03-01. F.L.V. was supported by CNPq-TWAS Sandwich Postgraduate Fellowship Award. O.L.G.P. acknowledges the support from grant 2012/16389-1, S$\tilde{\rm{a}}$o Paulo Research Foundation (FAPESP).

\bibliographystyle{aa}  
\bibliography{myrefs3}   

\end{document}